\documentclass[twocolumn,showpacs,preprintnumbers,amsmath,amssymb]{revtex4} 
\input{psfig.sty} 
\begin{document} 
 
\title{ 
Vertex cover problem studied by cavity method: 
Analytics and population dynamics 
} 
 
\author{ 
Haijun Zhou 
}                 
 
\affiliation{ 
Max-Planck-Institute of Colloids and Interfaces, 
14424 Potsdam, Germany 
} 
 
\date{February 14, 2003} 
 
\begin{abstract} 
We study the vertex cover problem on finite connectivity random graphs 
by zero-temperature cavity method. The minimum vertex cover 
corresponds to the ground state(s)  of a proposed Ising spin model. 
When the connectivity $c>e=2.718282$, there is no state for this  
system as the reweighting parameter $y$, which takes a similar 
role as the inverse temperature $\beta$ in conventional statistical 
physics, approaches infinity; 
consequently the ground state energy 
is obtained at a finite value of $y$ when the free energy function 
attains its maximum value. The minimum vertex cover size 
at given $c$ is estimated using population dynamics and compared with  
known rigorous bounds and numerical results. The backbone size  
is also calculated. 
\end{abstract} 
 
\maketitle 
 
\section{\label{sec.I} Introduction} 
The statistical physics of spin glass systems on infinitely connected 
lattices (e.g., the Sherrington-Kirkpatrick model) has been well  
understood \cite{mezard1987} and research interest is now focused on 
systems of finite connectivity (FC). For the limiting  example of  
FC random lattice, development of the cavity method has  
been made in recent few years \cite{mezard2001}. In a FC random lattice,  
each vertex interacts only with  a finite number of randomly chosen  
neighbours; and  the local  
structure of the lattice is tree-like, with the shortest 
distance between two randomly chosen vertices diverging as the 
system size goes to infinity.  This tree-like property makes it  
feasible to study the system by iterative cavity method \cite{note13.02.03}. 
The cavity method for $n$-dimensional regular lattice systems is 
yet to be worked out, and  some aspects of such kind of systems has been 
understood  using the well-developed mathematical tool of 
gauge transformation \cite{toulouse1977,nishimori2002}. 
 
The zero-temperature property of random lattice spin glasses is  
especially interesting. Here, the cavity formalism could be  
greatly simplified because only the minimum energy states 
are relevant \cite{mezard2003}.   
Furthermore, many combinatorial optimisation problems in 
computer sciences could be studied through a  mapping into an appropriate  
random spin glass model, examples of which include the 
K-sat \cite{monasson1997}, the XOR-sat \cite{ricci2001,caracciolo2002}, 
the vertex covering problem \cite{weigt2000,weigt2001}, the number partition  
problem \cite{mertens1998}, and so on.  
Recently, the zero-temperature cavity method was used on the random 
K-sat problem and the phase diagram for $k=3$ was obtained  
\cite{mezard2002,mezard2002b}. 
 
In this work we hope to improve our understanding on the vertex cover  
problem. Consider a random graph $G$ composed of $N$ vertices 
$V=\{1,2,\cdots,N\}$. Between any two vertices an edge is present 
with probability $c/(N-1)$ and absent with probability $1-c/(N-1)$, 
so that on average each vertex has $c$ neighbours  
(i.e., the average connectivity is $c$). The resulting edge set of 
graph $G$ is denoted by $E(G)$. A 
vertex cover of this graph consists of a set of vertices $\Lambda=\{i_1, 
i_2,\cdots, i_m\}$, with the property that 
if edge $(i,j)\in E(G)$, then either $i\in \Lambda$ or 
$j\in \Lambda$ or both \cite{weigt2000}. In the general case there  
are many different 
ways to cover a graph of size $N$; an interesting question is: 
Does there exist a vertex cover of size not exceeding $x N$ ($0<x<1$)? 
 
For large systems it was revealed that a sharp threshold value $x_c (c)$ 
exists. When $x>x_c(c)$, with probability approaching unity a  
vertex cover of size $\leq x N$ could be constructed for a given 
graph; while when $x< x_c (c)$ this probability approaches zero. 
This sharp threshold is also closely related to computational complexity.  
The vertex cover problem is NP-complete \cite{garey1979}, 
and a time growing exponentially with $N$ may be needed to determine 
whether a vertex cover of size $\leq x N$ exists or not for a given graph  
\cite{weigt2001b}. 
In the case of FC random graphs, when $x > x_c(c)$ or $x < x_c(c)$, it is 
relatively easy for a heuristic algorithm to check the existence of 
such a vertex cover; however, when $x \sim x_c(c)$, search 
complexity increases dramatically \cite{weigt2000}. 
For the practical purpose of designing better algorithms, it is 
important for us to understand the reason of this easy-hard-easy 
transition and to obtain a precise estimate of the threshold value 
$x_c(c)$. 
 
An rigorous bound exists for $x_c(c)$ \cite{gazmuri1984}: 
$x_l (c) < x_c(c) < 1-\ln c /c$, 
where $x_l (c)$ is the root of  
\begin{eqnarray} 
x_l (c) \ln x_l (c) +[1-x_l (c) ] \ln [1-x_l (c)] \nonumber \\
\;\;\;\;\;\;\;\;\;\;\;\;\;\;\;\;\;\;\;\; +(c/2) [1-x_l (c)]^2=0; 
\label{eq:gazmuri} 
\end{eqnarray} 
furthermore, $x_c (c)$ approaches the following asymptotic form at large $c$ 
\cite{frieze1990}: 
\begin{equation} 
\label{eq:frieze} 
x_c(c)=1-(2/c)[\ln c -\ln \ln c+1-\ln 2]+o(1/c). 
\end{equation} 
Using replica method of statistical physics, an analytical  
expression for $x_c$ was found in \cite{weigt2000}: 
\begin{equation} 
\label{eq:weigt} 
x_c(c)=1-W(c)/c-W^2(c)/ 2 c, 
\end{equation} 
where $W(c)$ is the Lambert-W-function defined by $W(c)\exp[W(c)]=c$. 
Equation (\ref{eq:weigt}) is exact when 
$c\leq e=2.718282$ \cite{bauer2001}. For $c >e$  
Eq.~(\ref{eq:weigt}) underestimates the true threshold value,  
and  for $c>20.7$ it is lower than the rigorous lower bound  
Eq.~(\ref{eq:gazmuri}). 
 
The present work focuses on the case of $c> e$. Using zero-temperature 
cavity method, we calculate both analytically and numerically the 
value of $x_c(c)$ and compare it with Eqs.~(\ref{eq:gazmuri}) and 
(\ref{eq:frieze}) and with numerical calculations reported in 
\cite{weigt2000}. In Section \ref{sec.II} an energy functional is  
introduced. In Section \ref{sec.III} the cavity formalism is 
outlined and the free energy expression is given. We investigate 
the $y\rightarrow \infty$ situation in Section \ref{sec.IV}  
($y$ is a reweighting parameter \cite{mezard2003,mezard2002,mezard2002b},  
it plays the role of the inverse 
temperature $\beta$ of conventional statistical mechanics). 
Section \ref{sec.V} reports the population dynamics results on 
the threshold value $x_c(c)$ and on the backbone size.  
We conclude the work in Section \ref{sec.VI}.

\section{\label{sec.II} The energy functional} 
We attach to  each vertex of the random graph $G$ an Ising spin  
$\sigma=\{-1,+1\}$. Associated with each spin micro-configuration  
is the following energy functional 
\begin{equation} 
\label{eq:energy} 
E[\{\sigma_i\}]=-\sum\limits_{i=1}^N \sigma_i+{\lambda \over 2}  
\sum\limits_{(i,j)\in E(G)} (1+\sigma_i)(1+\sigma_j), 
\end{equation} 
where $\lambda$ is an constant parameter chosen to be greater than unity  
\cite{note11.11.02}. 
Denote $N_{\rm mvc}(G)$ as the size of the minimum vertex 
cover(s) of graph $G$,  then the minimum energy  
over all the $2^N$ possible spin  configurations of graph $G$ is 
\begin{equation}	 
\label{eq:min_energy}	 
E_{\rm min}=2 N_{\rm mvc} (G) -N. 
\end{equation} 
Some explanation on Eq.~(\ref{eq:min_energy}). First, $E_{\rm min}$ is 
reachable. We denote $\Lambda_{\rm mvc}(G)$ as (one of) the  
minimum-sized vertex cover(s), and assign $\sigma=-1$ to vertices in 
$\Lambda_{\rm mvc} (G)$ and $\sigma=+1$ to vertices outside. The 
energy of this spin configuration is $2 N_{\rm mcv} (G)-N$. 
Second, no spin configuration could have lower energy. 
To see this, suppose that another spin configuration has lower energy  
than Eq.~(\ref{eq:min_energy}). This 
spin configuration must contain $N_a <N_{\rm mvc} (G)$ negative spins, 
with $N_a+\lambda N_0 < N_{\rm mvc} (G)$, where  
$N_0=(1/4)\sum_{(i,j)\in E(G)}(1+\sigma_i) (1+\sigma_j)$. 
However, one at most need to change the spin values of $N_0$ vertices  
from $+1$ to $-1$ to make 
the sum $\sum_{(i,j)\in E(G)}(1+\sigma_i)(1+\sigma_j) =0$, and 
the resulting new set of negative spin vertices is a 
vertex cover with size $N_a+N_0 <N_{\rm mvc}(G)$. This conflicts with  
our original assumption that $N_{\rm mvc} (G)$ is  
the minimum vertex cover size.  
 
The problem of finding $x_c (c)$  is converted to finding the average of  
$E_{\rm min}/N$ over the random graphs $G$: 
\begin{equation} 
\label{eq:x_c_cvt} 
x_c(c)=(1/2)(1+\overline{E_{\rm min}}/N). 
\end{equation} 
Here $\overline{(\cdot)}$ means the average of $(\cdot)$ over different 
realizations of the random graph. We use cavity method to estimate  
$\overline{E_{\rm min}}/N$. In the next section, 
the cavity formalism for the present problem is outlined.  
The reader is referred to  
\cite{mezard1987,mezard2001,mezard2003,mezard2002,mezard2002b} for 
more detailed discussion.  
 
\section{\label{sec.III} The zero-temperature cavity formalism} 
 
At zero temperature, only the minimum energy configurations are  
relevant. There could be a great many energy local-minima for 
Eq.~(\ref{eq:energy}). For very large system size $N$, we group 
these configurations into different $``$states''. A state of the 
system corresponds to a set of spin micro-configurations. These 
spin configurations all have the same energy, which is a 
local minimum of Eq.~(\ref{eq:energy}); and two such spin configurations 
differ only in a finite number of spin flips. 
The average number of states at given density $\epsilon$ of  
local minimum energy is assumed to be an exponentially increasing 
function of system size $N$, and is characterised by the entropy 
density $\Sigma(c,\epsilon)$. We can introduce a reweighting parameter 
$y$ and define an zero-temperature free energy density $\Phi(y)$ 
through the following equation 
\begin{equation} 
\label{eq:free_energy} 
\int d\epsilon \exp[-N y \epsilon+ N \Sigma(c,\epsilon)]  
=\exp[-N y\Phi(y)]. 
\end{equation} 
Equation (\ref{eq:free_energy}) has the same form as the conventional 
definition of free energy in  textbooks of equilibrium statistical 
physics. The reweighting parameter $y$ plays the role of inverse 
temperature. A large value of $y$ ensures that states with  
lower energies will be favoured, provided that such states 
exist (i.e., $\Sigma(c,\epsilon)\geq 0$). 
 
Suppose we have a system of $N$ vertices (spins).  Now  add a 
new spin $\sigma_0$ into the system and connect it to $k$ preexisting 
spins $\sigma_1,\cdots,\sigma_k$, where $k$ obeys the Poisson distribution 
of mean $c$, $P_P(k,c)=e^{-c} c^k /k!$. The energy of the $N$-spin system  
at fixed value of the spins $\sigma_1,\cdots,\sigma_k$ is supposed to be 
\begin{equation} 
\label{eq:e_n} 
E^{(N)} (\sigma_1,\cdots,\sigma_k)=A-\sum\limits_{i=1}^k h_i \sigma_i, 
\end{equation} 
with $A$ being a constant.  
In the above equation, $h_i$  is the  
local field (called the cavity field) felt by spin $\sigma_i$ in the 
absence of $\sigma_0$ in a given macroscopic state. Since the graph is  
locally tree-like, as the system size becomes very large, the shortest distance  
between two randomly chosen cavity spins also becomes large; therefore, 
the cavity field $h_i$ felt by spin $\sigma_i$ becomes independent 
of the values of all the other cavity spins \cite{note13.02.03b}.  
After the addition of spin $\sigma_0$, the 
minimum energy of the $(N+1)$-spin system at fixed value of $\sigma_0$ is 
\begin{equation} 
\label{eq:e_n+1} 
E(\sigma_0)=\left\{\begin{array}{ll} 
A-\sigma_0, &\;\;{\rm if}\;\;k=0 \\ 
A-\sum_{i=1}^k \hat{w}(h_i) & \; \\
\;\;\;\; -[1+\sum_{i=1}^k  
\hat{u}(h_i)] \sigma_0, &\;\;{\rm if} 
\;\;k\geq 1. 
\end{array} 
\right. 
\end{equation} 
Here,  
\begin{equation} 
\label{eq:wu} 
\begin{array}{llll} 
\hat{w}(h) =  0  & \;\;{\rm if}\;\;h=1 &\;\;\;{\rm and} 
\;\;=|h| &\;\;{\rm if}\;\;{h\leq 0}, \\  
\hat{u}(h) =  -1 &\;\;{\rm if}\;\;h=1 &\;\;\;{\rm and}\;\;=0 
&\;\;{\rm if}\;\;{h\leq 0}. 
\end{array} 
\end{equation} 
(We have used the fact  that the cavity fields at zero temperature are  
integer-valued and do not exceed unity \cite{note13.02.03b}.) The energy 
shift caused by the addition of spin $\sigma_0$ is  
$\triangle E_1=-1$ (if $k=0$) and $\triangle E_1=-\sum_{i=1}^k  
[\hat{w}(h_i)-|h_i|]-|1+\sum_{i=1}^k \hat{u}(h_i)|$ (if $k\geq 1$). 
 
Equation (\ref{eq:e_n+1}) indicates that the cavity field at 
spin $\sigma_0$ is $h_0=1$ (if $k=0$) or $h_0=1+\sum_{i=1}^k \hat{u}(h_i)$ 
(if $k\geq 1$).  By definition \cite{note13.02.03b}, the cavity field 
of a spin at each state has unique value; however, its value may 
be different for different states. Denote the probability distribution of  
the cavity field at vertex $i$ among different states as 
$P_i(h)$ (it is called  the h-survey  in Ref.~\cite{mezard2002b}). 
Because different vertices have different local 
environments, the h-surveys are different for different vertices.  
With the introduction of the reweighting parameter $y$ which favours 
lower-energy states, the h-survey at spin $\sigma_0$  
is related to those of the cavity spins by 
\begin{widetext}
\begin{equation} 
\label{eq:h_survey} 
P_0(h)=\delta_k^0 \delta(h-1)+ 
[1-\delta_k^0] C\int \prod\limits_{i=1}^k [P_i(h_i) d h_i ] 
\delta [h-1-\sum\limits_{i=1}^k \hat{u}(h_i)] \exp(-y \triangle E_1), 
\end{equation}
where $C$ is an normalisation constant.  
Eq.~(\ref{eq:h_survey}) is actually a self-consistent equation 
for the h-survey.  A careful analysis of Eq.~(\ref{eq:h_survey}) leads to 
the following expression 
\begin{equation} 
\label{eq:ph_form} 
P(h)=\left\{ 
\begin{array}{ll} 
\sum_{l=0}^\infty \zeta_l \delta(h+l),  
& \;\;\;\;\;\;{\rm with}\;\;{\rm probability}\;\;p_1 \\ 
\delta(h-1), & \;\;\;\;\;\;{\rm with}\;\;{\rm probability}\;\;p_2 \\ 
\alpha \sum_{l=0}^\infty \zeta_l \delta(h+l) +(1-\alpha) \delta(h-1), &  
\;\;\;\;\;\;{\rm with}\;\;{\rm probability}\;\;p_3=1-p_1-p_2  
\end{array}\right. 
\end{equation} 
\end{widetext} 
where $\zeta_l\geq 0$ and $\sum_l \zeta_l=1$;  
$\alpha\in (0,1)$ is determined by certain probability distribution 
$\rho(\alpha)$, and 
\begin{eqnarray} 
\label{eq:p1p2} 
p_1&=& 1-\exp(-c p_2), \nonumber \\ 
p_2&=& \exp(-c (1-p_1)). 
\end{eqnarray} 
 
For very large system size $N$, Eq.~(\ref{eq:free_energy}) suggests 
that at fixed value of $y$,  
$\Phi(y)=\epsilon-\Sigma(c,\epsilon)/y$, with $\epsilon$ being 
implicitly determined by $\partial \Sigma(c,\epsilon)/\partial \epsilon 
=y$. An explicit expression for the free energy density could be 
obtained by the following way: 
 
After the addition of spin $\sigma_0$, the averaged number of states  
of the $(N+1)$-system is 
\begin{widetext}
\begin{equation} 
\exp[(N+1) \Sigma({2 E(G(N))+2 k \over N+1},{E \over N+1})] 
=
\int d \triangle E_1 P^{(1)}(\triangle E_1)  
\exp[N \Sigma({2 E(G(N))\over N}, {E-\triangle E_1 \over N})], 
\label{eq:entropy} 
\end{equation} 
where $P^{(1)}(\triangle E_1)$ is the probability distribution 
function of the energy shift $\triangle E_1$: 
\begin{equation} 
\label{eq:p_add} 
P^{(1)}(\triangle E_1)= 
\delta_k^0 \delta(\triangle E_1 +1) 
+(1-\delta_k^0) \int  
\prod\limits_{i=1}^k [P_i(h_i) d h_i]  
\delta[\triangle E_1+\sum\limits_{i=1}^k (\hat{w}(h_i)-|h_i|) 
+|1+\sum\limits_{i=1}^k \hat{u}(h_i)|]. 
\end{equation} 
A logarithm operation is performed on Eq.~(\ref{eq:entropy}), 
and the resulting equation is averaged over all the  
possible realizations of the cavity fields 
and $k$ (this average operation is denoted by an overbar in the 
following equation).  We obtain that 
\begin{equation} 
-y \Phi(y)= 
\Sigma(c,\epsilon)-y\epsilon
=-c {\partial \Sigma(c,\epsilon) \over \partial c} 
+\overline{ 
\ln \int d \triangle E_1 P^{(1)} (\triangle E_1)  
\exp(-y \triangle E_1)}. 
\label{eq:add} 
\end{equation}	 
To compute $\partial \Sigma/\partial c$, we setup an edge between 
two cavity spins. The energy shift, $\triangle E_2$,  
caused by this new edge obeys the following distribution 
\begin{equation} 
P^{(2)}(\triangle E_2)=\int \prod\limits_{i=1}^2 [d h_i  P_i(h_i)] 
\delta[\triangle E_2-({\rm min}_{\sigma_1,\sigma_2} 
[{\lambda \over 2}(1+\sigma_1)(1+\sigma_2)-h_1 \sigma_1 -h_2 \sigma_2] 
+|h_1|+|h_2|)]. 
\end{equation} 
The averaged number of states of the new system is 
\begin{equation} 
\exp[N \Sigma({2E(G(N))+2 \over N},{E \over N})]= 
\int d \triangle E_2 P^{(2)}(\triangle E_2) 
\exp[N \Sigma({2 E(G(N)) \over N}, {E -\triangle E_2 \over N})]. 
\end{equation} 
After performing the same procedure as mentioned below 
Eq.~(\ref{eq:p_add}),  we find that 
\begin{equation} 
\label{eq:link} 
{\partial \Sigma(c,\epsilon) \over \partial c} 
= 
(1/2)\overline{ 
\ln \int d \triangle E_2 P^{(2)}(\triangle E_2) 
\exp(-y \triangle E_2)}. 
\end{equation} 
The free energy density expression could be obtained from 
Eqs.~(\ref{eq:add}) and (\ref{eq:link}). After taking into 
consideration Eq.~(\ref{eq:ph_form}), we arrive at the 
following expression for the free energy density: 
\begin{eqnarray} 
\label{eq:fe_1} 
\Phi(y)&=&2 p_1-1-c p_2^2+{c p_2 p_3 \over y} \int d \alpha  
\rho(\alpha) \ln (\alpha+(1-\alpha) e^{-2 y})  \nonumber \\ 
 & &-{p_3 \over y} \sum\limits_{m=1}^\infty  
{P_P(m,c p_3) \over 1-e^{-c h}} 
\int \prod\limits_{i=1}^m [\rho (\alpha_i) d \alpha_i] 
\ln (e^{-2 y} +(1-e^{-2 y})\prod\limits_{i=1}^m \alpha_i) \nonumber \\ 
& & 
+{c p_3^2 \over 2 y} \int \prod\limits_{i=1}^2 [ d \alpha_i  \rho(\alpha_i)] 
\ln (1-(1-e^{-2 y}) \prod\limits_{i=1}^2 (1-\alpha_i)). 
\end{eqnarray} 
 
At given $y$, the energy density and the entropy density are calculated by 
$\epsilon = \Phi(y)-y{d \Phi(y)/d y}$ and  
$\Sigma = y^2 {d \Phi(y)/ d y}$, respectively. 
\end{widetext}
 
\section{\label{sec.IV} The $y\rightarrow \infty$ limit} 
 
At this stage, it is helpful for us to introduce an auxiliary probability 
distribution function called the u-survey \cite{mezard2002b}: 
\begin{equation} 
Q_i(u)=C\int d h P (h) \delta(u-\hat{u}(h)) \exp(y (\hat{w}(h)-|h|)). 
\end{equation} 
Based on Eq.~(\ref{eq:ph_form}) we know that 
$Q_i(u)=\delta(u)$ (with probability $p_1$) or $Q_i(u)= 
\delta(u+1)$ (with probability $p_2$), or 
\begin{equation} 
\label{eq:u_hybrid} 
Q_i(u)=\eta \delta(u)+ [1-\eta] \delta(u+1). 
\end{equation} 
The hybrid h-survey and u-survey at a given vertex are related by 
$\alpha=\eta/(\eta+(1-\eta) e^y)$. The distribution of the $\eta$  
value in the hybrid u-survey Eq.~(\ref{eq:u_hybrid}) is 
governed by 
\begin{widetext} 
\begin{equation} 
\label{eq:eta} 
\rho(\eta)=\sum\limits_{m=1}^{\infty} {P_P(m,c p_3) \over 
1-e^{-c p_3}} \int \prod\limits_{i=1}^n  [\rho(\eta_i) d \eta_i ] 
\delta(\eta-1+{e^{y} \prod\limits_{l=1}^n \eta_l \over 
\prod\limits_{l=1}^n [\eta_l+e^{y} (1-\eta_l)]+(e^{y}-1)  
\prod\limits_{l=1}^n \eta_l}). 
\end{equation} 
The free energy expression of Eq.~(\ref{eq:fe_1}) 
could be rewritten in the following form 
\begin{eqnarray} 
\label{eq:fe_2} 
\Phi(y) &=& p_1-p_2-{c\over 2} [(1-p_1)^2+(p_2)^2]  
 +{p_3 (1+c p_2) \over y} \overline{\ln [\eta+e^{-y} (1-\eta)]} \nonumber \\ 
 &-& {p_3 [1+c(1-p_1)] \over y} \overline{\ln [1-\eta+e^{-y} \eta]}  
+{c p_3^2 \over 2 y} \overline{\ln [e^{-y}+(1-e^{-y}) 
(\eta_1+\eta_2-2 \eta_1 \eta_2)]}. 
\end{eqnarray} 
\end{widetext}
The overbars in Eq.~(\ref{eq:fe_2}) denote 
the average over the $\eta$ distribution given by 
Eq.~(\ref{eq:eta}). 
 
When $c\leq e$, the only solution of Eq.~(\ref{eq:p1p2}) is that 
$p_1=1-p_2$, $p_2=\exp(-c p_2)=W(c)/c$, and $p_3=0$. In this case we  
recover Eq.~(\ref{eq:weigt}). $\Sigma(c,\epsilon) =0$ for this solution, 
indicating that there is only one state. 
It was found that for $c\leq e$, all the vertices of the random graph 
could be removed by application of an leaf-removal algorithm \cite{bauer2001}.  
 
When $c >e$ the above-mentioned solution becomes unstable,  
and a new solution appears with $p_2=\exp[-c \exp(-c p_2)]  
<\exp(-c p_2)$, $p_1=1-\exp(-c p_2)$, and $p_3 >0$. The analysis of 
Bauer and Golinelli \cite{bauer2001} reveals that a core with size 
proportional to $N$ remains after application of the  
leaf-removal algorithm. The core is a strongly connected subgraph, in which 
each vertex is connected  at least to two other vertices. 
 
To estimate the value of $x_c(c)$ for $c >e$,  
let we first consider the limiting situation of $y\rightarrow \infty$.  
Equation (\ref{eq:free_energy}) ensures  that the minimum $\epsilon$  
corresponds to $y=\infty$, provided that the configurational entropy  
is  non-negative at this limit. 

\begin{figure} 
\psfig{file=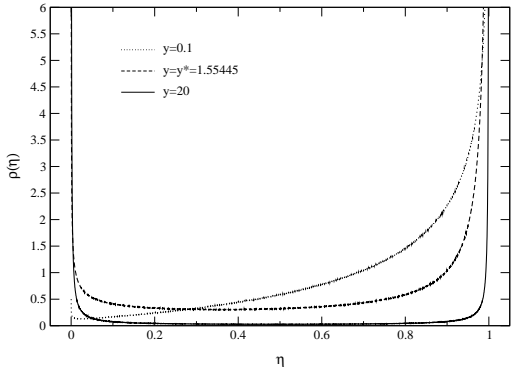,angle=270,width=1.\linewidth}
\caption{\label{fig:fig01}  
The $\eta$ value distribution Eq.~(\ref{eq:eta}) obtained 
by population dynamics at $c=6.0$. The $\rho(\eta)$ distribution 
strongly depends on the value of the reweighting parameter $y$.  
In this figure, three curves at $y=0.1$ (dotted line),  
$y=y^*=1.5545$ (corresponding to the free energy maximum, dashed 
line), and $y=20$ (solid line) are shown.} 
\end{figure}

From Eq.~(\ref{eq:eta}) we know that as $y\rightarrow \infty$,   
\begin{equation} 
\label{eq:eta_limit} 
\rho(\eta)=r_1 \delta(\eta-0^+) +r_2 \delta(\eta-1^-)+r_3 \rho^* (\eta), 
\end{equation} 
where $\rho^* (\eta)$ is a uniform distribution 
over $(0,1)$. Equation~(\ref{eq:eta_limit}) is confirmed 
by the population dynamics calculation (Fig.~\ref{fig:fig01}).  
It is easy to verify that 
\begin{equation} 
\label{rrr} 
r_1={1-c p_2 \over c p_3},\;\;\;r_2={c+c p_1-\ln c \over c p_3},\;\;\; 
r_3={\ln c-1 \over c p_3}. 
\end{equation} 
Consequently, we obtain from Eq.~(\ref{eq:fe_2}) that 
\begin{eqnarray} 
\label{eq:1rsb} 
\epsilon_{\infty} & = & 1-{\ln^2(c) /2 c}-{\ln (c) / c} 
-{3 / 2 c}, \nonumber \\ 
\Sigma_\infty & = & -{\pi^2 (\ln c -1)^2 / 16 c}. 
\end{eqnarray} 
The minimum energy  value given in the above equation  
is an improved lower-bound 
of the true average minimum energy. It was obtained also in 
\cite{weigt2001} by replica method. However, at $y=\infty$ 
the entropy density $\Sigma_\infty$ is negative, suggesting that there is no  
state at this energy density. Indeed the $x_c(c)$ value  
[Eq.~(\ref{eq:x_c_cvt})] of this solution 
also exceeds the rigorous lower-bound when $c>27.3$.  
The true energy density must be higher than the value  
given in Eq.~(\ref{eq:1rsb}). 
A better estimate of the minimum energy density could be 
obtained by calculating  the maximum value of $\Phi(y)$ with respect 
to the reweighting parameter $y$ \cite{note11.11.02b}. 
This is done in the next  
section with population dynamics \cite{mezard2001}. 
 
\section{\label{sec.V} Population dynamics at finite $y$} 
 
\begin{figure} 
\psfig{file=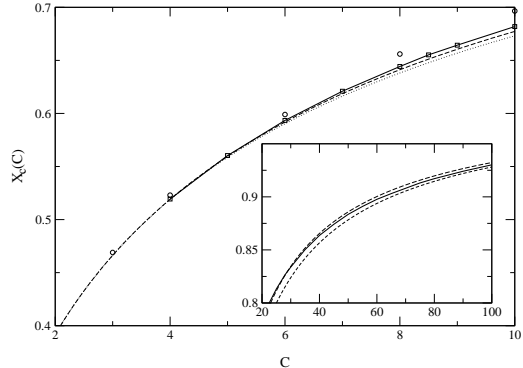,angle=270,width=1.0\linewidth}
\caption{\label{fig:fig02} 
The minimum vertex cover size  
$x_c(c)$ estimated by population dynamics (solid line and squares)  
and its comparison with the exact numerical enumeration values of 
\cite{weigt2000} (circles), the replica-symmetric estimation  
Eq.~(\ref{eq:weigt}) (dotted line), and the 
estimation given by Eq.~(\ref{eq:1rsb}) (dashed line). 
In the inset, the $x_c(c)$ value obtained by population dynamics 
(solid line) is compared with the rigorous lower-bound Eq.~(\ref{eq:gazmuri}) 
(dashed line) and the asymptotic value Eq.~(\ref{eq:frieze}) (long dashed 
line). 
} 
\end{figure}

To obtain the value of $\Phi(y)$ at any given $y$, the technique of  
population dynamics is used \cite{mezard2001}.  
A large population of $\eta$'s is generated and this population then  
evolves according to Eq.~(\ref{eq:eta}). The averages in  
Eq.~(\ref{eq:fe_2}) are calculated numerically. The resulting 
estimates of $x_c(c)$ are shown in Fig.~\ref{fig:fig02} and 
listed in Table~\ref{tab:tab01}.  

\begin{table} 
\caption{\label{tab:tab01} 
The $x_c(c)$ value obtained by the population dynamics (column 2)  
and its comparison with the exact numerical enumeration value 
reported in \cite{weigt2000} (column 4). 
The reweighting parameter value $y^*$ of column 3 at 
given $c$ corresponds to the maximum of the free energy density  
Eq.~(\ref{eq:fe_2}). To determine $y^*$,  
Eq.~(\ref{eq:fe_2}) is fitted with two adjustable parameters: 
 $\Phi(y)=\epsilon_\infty+C_1 /y+C_2 \exp(-y)/y$, where 
$\epsilon_\infty$ is given in Eq.~(\ref{eq:1rsb}). The  
population size adopted in the present work is $20,000$. 
} 
\begin{tabular}{cccc} 
$c$	& $x_c$ (cavity) & $y^*$ &$x_c$ (enumeration) \\ \hline 
$4.0$ 	& $.5194$	& $1.3093$	& $.523\pm .001$ \\ 
$5.0$	& $.5603$	& $1.5758$	& \\ 
$6.0$	& $.5934$	& $1.5545$	& $.599\pm .001$ \\ 
$7.0$	& $.6210$	& $1.5260$	& \\ 
$8.0$	& $.6443$	& $1.5259$	& $.656\pm .003$ \\ 
$9.0$	& $.6643$	& $1.5326$	& \\ 
$10.0$	& $.6819$	& $1.5434$	& $.697\pm .003$ \\ \hline 
\end{tabular} 
\end{table}

The threshold curve $x_c (c)$ obtained by the present method lies within the 
rigorous bound given by Eq.~(\ref{eq:gazmuri}). It is therefore an improved 
estimate compared with Eq.~(\ref{eq:weigt}) and Eq.~(\ref{eq:1rsb}), both 
of which exceed the rigorous lower-bound when $c$ is larger than certain  
value. However, the values of $x_c$ estimated by the cavity method is 
systematically smaller than enumeration results on  
finite systems (Fig.~\ref{fig:fig02}). 
When the average connectivity $c$ is large, 
the estimated $x_c(c)$ value approaches the asymptotic value but 
it lies  below the asymptotic curve (Fig.~\ref{fig:fig02} inset). 
These discrepancies suggest that the $x_c(c)$ threshold value 
obtained by the cavity method is not exact; it could 
serve as an improved lower-bound of the real threshold  curve. 
The reason for the failure of the cavity method to obtain exact 
threshold values for $c>e$ are discussed in the next section.

\begin{figure} 
\psfig{file=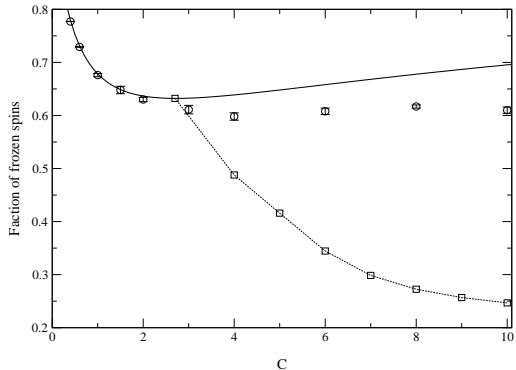,angle=270,width=1.0\linewidth}
\caption{\label{fig:fig03} 
Fraction of frozen spins calculated by population dynamics  
\cite{note11.11.02c} (squares) and 
its comparison with numerical result reported in \cite{weigt2001} and 
the replica-symmetric result \cite{weigt2001} (thick solid line). 
} 
\end{figure} 
 
The cavity field distribution Eq.~(\ref{eq:ph_form}) contains information 
on the spin state at a given vertex. A spin $\sigma_i$ will be fixed  
at $\sigma_i=+1$ if the local cavity field is distributed according to 
$P(h_i)=\delta(h-1)$ or $P(h_i)=\alpha \sum_{l=0}^\infty 
\zeta_l \delta(h+l)+(1-\alpha)\delta(h-1)$ but with  
$\alpha\rightarrow 0^{+}$ \cite{note11.11.02c}; on the other hand, $\sigma_i$ will 
be fixed at $\sigma_i=-1$ with probability $\sum_{l=1}^\infty \zeta_l$ 
if $P(h_i)=\sum_{l=0}^\infty \zeta_l \delta(h+l)$ or  
$P(h_i)=\alpha \sum_{l=0}^\infty \zeta_l \delta(h+l)+(1-\alpha)\delta(h-1)$  
but with $\alpha\rightarrow 1^-$ \cite{note11.11.02c}. For the minimum vertex covers 
at $y=y^*$, the probability for a randomly 
chosen vertex to have fixed spin value is calculated by 
population dynamics (see Fig.~\ref{fig:fig03}). 
The fraction of frozen spins calculated by the population dynamics 
is much lower that that obtained by numerical enumeration \cite{weigt2001}. 
This discrepancy may be further indication that the full hierarchy of 
replica symmetry breaking is needed to completely describe the 
property of the minimum vertex covers. We have noticed  
that the fraction of frozen spins strongly depends 
on the value of the reweighting parameter $y$. When $y\rightarrow \infty$ 
the replica-symmetric result of \cite{weigt2001} is recovered; 
while for $y\sim 2 y^*$--$3 y^*$ the data obtained by numerical 
enumeration is approached.  
 
\section{\label{sec.VI} Discussion} 
 
In this work, we estimate the average minimum vertex cover size $x_c(c)$   
for random graphs of finite connectivity $c>e$ in the large $N$ limit.  
The obtained $x_c(c)$ curve lies within the rigorous bound 
\cite{gazmuri1984} and approaches the asymptotic curve 
Eq.~(\ref{eq:frieze}) at large $c$ values. It could be regarded as an  
improved lower-bound of the real threshold $x_c(c)$ value. 
 
When $c>e$ the threshold $x_c$ estimated by the cavity method is 
not exact. The reason may be the following:  The cavity method  
is equivalent to one-step replica-symmetry-breaking 
\cite{mezard1987,mezard2001,mezard2003}, and the cavity 
fields on different vertices are considered as uncorrelated. Because 
of the core percolation beyond $c=e$ \cite{bauer2001} in the random 
graph, the cavity fields of different vertices may actually be 
correlated strongly. To partly account for this effect, one possibility 
is to consider also non-integer cavity fields. We hope to return to 
this point in a latter work. 
 
The cavity method has inspired  very efficient algorithms to 
tackle the random K-sat problem.  
In the random K-sat problem, there 
exists a glassy phase at $y\rightarrow \infty$ for certain range of 
the connectivity $c$ \cite{mezard2002,mezard2002b}. In this glassy phase, 
the minimum energy is still located at $y=\infty$ but the  
complexity is positive, and the algorithm based on 
the idea of cavity field \cite{mezard2002b} works 
well in this phase. For the vertex cover problem, the present 
work suggests that such a $y=\infty$ glass phase does not exist 
(this conclusion seems also true for the vertex cover problem on 
random hyper-graphs where each $''$edge" is a triangle \cite{weigt2003}). 
This indicates that vertex covers of size $N x_c(c)+O(1)$ are extremely 
few. It remains to 
be seen whether or not algorithms based on cavity method could efficiently  
find vertex covers of size slightly beyond $N x_c(c)$ for a given  
random graph. 
 
\section*{Acknowledgment} 
This work is made possible by a post-doctoral fellowship of 
the Max Planck Society. 
I thank M. Weigt for suggestions concerned with earlier 
versions of this manuscript, and  A. K. Hartmann and M. Weigt 
for sharing numerical data. I also thank Professor R. Lipowsky and 
professor  YU Lu for support. 
 

\begin{thebibliography}{24}
\bibitem[{not({\natexlab{b}})}]{note11.11.02}
\bibinfo{note}{The value of $\lambda$ is irrelevant as long as $\lambda > 1$:
  The bond energy, $(\lambda/2)(1+\sigma_i)(1+\sigma_j)$, equals $0$ or
  $2\lambda$; if it is $2 \lambda$, one can choose either $\sigma_i=-1$ or
  $\sigma_j=-1$ and the total energy will be lowered. However, choosing
  $\lambda\simeq 1$ might facilitate a search algorithm in finding the true
  energy minimum.}
\bibitem[{not({\natexlab{c}})}]{note13.02.03b}
\bibinfo{note}{Some more discussion about the cavity fields. At zero
  temperature, each state of the system corresponds to an ensemble of spin
  microscopic configurations. All these configurations have the same energy
  (corresponding to a minimum value of the energy functional
  Eq.~(\ref{eq:energy})), and only finite number of spin flips is needed to
  transit from one to another configuration of them. The spins on some of the
  vertices are fixed at $\sigma=+1$ or $\sigma=-1$ in all these configurations,
  while the spins on the remaining vertices are not fixed. We say that those
  spins whose value fluctuates from one configuration to another do not feel
  any magnetic field in this state ($h=0$). A spin whose value is fixed feels a
  non-zero magnetic field. The magnitude of this non-zero cavity field is
  defined as half the total increase in minimum energy associated with flipping
  this spin. For our present problem, an energy local minimum has value $2
  N_{\downarrow} -N$, where $N_{\downarrow}$ is the total number of negatively
  valued spins. Therefore, we conclude that (1) all the cavity fields must be
  integer-valued, and (2) the cavity fields could not exceed unity ($h\leq
  1$).}
\bibitem[{not({\natexlab{d}})}]{note11.11.02b}
\bibinfo{note}{This is understood by the following consideration: According to
  Eq.~(\ref{eq:free_energy}), to obtain the energy absolute minimum the
  reweighting parameter $y$ should be set to its largest possible value. On the
  other hand, $\Sigma=y^2 d\Phi (y)/d y$ changes sign from positive to negative
  as $y$ is increased beyond some value. Therefore the largest $y$ value is
  obtained by the equation $d \Phi(y)/ d y=0$, which corresponds to the maximum
  of $\Phi(y)$.}
\bibitem[{not({\natexlab{e}})}]{note11.11.02c}
\bibinfo{note}{In the numerical calculation, $\alpha\rightarrow 0^+$ is
  replaced by the condition $\alpha\leq \alpha_c$ and $\alpha\rightarrow 1^-$
  is replaced by $\alpha\geq 1-\alpha_c$ with $\alpha_c$ chosen to be $0.01$.
  It turns out that the calculated fraction of frozen spins is sensitive to the
  $\alpha_c$ value. A more stringent criterion (say $\alpha_c=0.001$) leads to
  even smaller fraction, while a looser criterion (say $\alpha_c=0.05$) leads
  to much bigger value.}
\end{thebibliography}

\end{document}